\documentclass[reprint,superscriptaddress,twocolumn,showpacs,amsmath,amssymb]{revtex4}

\usepackage{graphicx}

\newcommand{\bj}{\mathbf{j}} \newcommand{\bV}{\mathbf{V}}

\begin{document}

\title{Distinguishability of particles and its implications \\ for
  peculiar mass transport in inhomogeneous media}


\author{Sergei~P.~Lukyanets} \email[Email address:
]{lukyan@iop.kiev.ua} \affiliation{Institute of Physics of National
  Academy of Sciences of Ukraine, Prospect Nauki 46, 03028 Ky\"{\i}v,
  Ukraine}

\author{Oleksandr~V.~Kliushnychenko} \affiliation{Department of
Natural Sciences, National University of Ky\"{\i}v-Mohyla Academy, 2
Skovorody Str., 04655 Ky\"{\i}v, Ukraine }

\begin{abstract}
A mass transfer directed from low to high density region in
an inhomogeneous medium is
modeled as a limiting case of a two-component lattice gas with excluded
volume constraint and one of the components fixed. In the long-wavelength
approximation, density relaxation of mobile particles is governed by a
diffusion process and interaction with a medium inhomogeneity represented by
a static component distribution. It is shown that density relaxation can
be locally accompanied by a density distribution compression. In
quasi one-dimensional case, the compression dynamics manifests itself in a hopping-like motion of
diffusing substance packet front position due to a staged passing
through inhomogeneity barriers and leads
to a fragmentation of a packet and retardation of its spreading.
A
root-mean-square displacement reflects only an averaged packet front
dynamics and becomes unappropriate as a transport characteristic in
this regime.
In a stationary case mass transport throughout
a whole system may be directed from a boundary with low concentration
towards a boundary with that of high one. Implications of the excluded
volume constraint and particles distinguishability for
these effects are discussed.

\end{abstract}

 \maketitle

\section{\label{intro}Introduction}

Mass transfer in inhomogeneous media and complex systems is often
governed by nonclassical diffusion laws and exhibits peculiar
transport effects. Specific examples include sub- or superdiffusion
processes of, for example, Brownian particles in hydrodynamic flow
\cite{1,2,3,B4}, and absolute negative mobility effect
\cite{N1,N2,N3chan,N4chan,N5chan,N6,N7int,N8int,N9int}. Very often,
studying such effects we encounter a situation where relaxation goes
in counterintuitive way, namely, instead of expected spreading density
distribution undergoes compression, at least over certain time period,
or even full collapse. In other words, mass transport is directed from
low to high density region. In particular, such behavior occurs in
systems with absolute negative mobility or negative diffusion
coefficient.

The phenomenon of absolute negative mobility or negative diffusion
coefficient may be of different origin. In particular, it can occur
due to time- or particle-particle correlations, or nonequilibrium
character of a process. This effect was demonstrated, for example, in
electron-hole plasma, as an electron-drag effect arising from an
electron-hole scattering \cite{N1}, in models of interacting Brownian
particles \cite{N7int,N8int,N9int} or interacting lattice gas
\cite{N11Chum}, and for nonequilibrium dynamics of Brownian particle
in a periodic one-dimensional potential or a quasi one-dimensional channel
in a presence of external force periodic in time
\cite{N3chan,N4chan,N5chan}. Note that the negative sign of diffusion
coefficient entails absolute instability in a system and may signify
the onset of a new phase formation, e.g., nuclei growth at first order
phase transition \cite{N11Chum, N12} or collapse of electron-hole
plasma that may be a precursor for a neutral exciton formation.
\cite{N10efr}.

The phenomenon of mass transfer from low to high density region also
may be caused by an interaction of particles with medium
inhomogeneity. In this case density relaxation through this
interaction may go much faster than via diffusion process, and, as a
result, the latter may not be a main relaxation and entropy production
mechanism \cite{N10efr}. A simple, although speculative, example of
such a density relaxation is a non-wettability effect, caused by
interaction between a liquid drop and a substrate (medium). Analogous
effect in diffusive systems will be of primary interest in this paper.

Here we demonstrate that such an effect may naturally appear for
diffusing particles in inhomogeneous media. In this case density
relaxation dynamics in a subdiffusive regime can be locally
accompanied by compression of density distribution. Moreover, in a
stationary case, mass transport through the inhomogeneous sample can
go from a boundary with low concentration to a boundary with that of
high one.

In order to demonstrate these effects we resort to the simplest model
of a two-component lattice gas assuming that every lattice site can be
occupied by one particle only (excluded volume constraint). In the
limiting case of a frozen or static component this model reduces to the
mass transport of mobile component in the inhomogeneous
medium. Another reason to use such a model is an appearance of a drag
effects, which has been found in this framework in a presence of
driving field and nonzero particle hopping rates of both components,
see, e.g., \cite{S2,S3}. In addition, absolute negative mobility
effect may appear in a case of non-Markov dynamics \cite{N6}. Two
component lattice gas may also display memory effects.  For example,
correlation between jumps of tagged particle and its influence on
tracer diffusion was examined in \cite{N13,Chumak2002}. This
correlation represents the tendency of a tagged particle to return to
its previous position and it follows from: (i) a presence of different
particle sort, i.e. particle distinguishability, and (ii) the
excluded volume constraint. The latter corresponds to infinite
repulsion energy between particles at the same lattice site. Formally,
such ``back correlations'' describe a repulsion of a tagged atom from
the other gas component.

\section{\label{model} The Model}

We consider the simplest model of a two-component lattice gas on a
regular lattice, see, e.g., \cite{N13}. It is assumed that every
lattice site can be occupied by a single particle of a sort $m$ or $n$
only. Particles can hop between neighboring sites separated by
a distance $a$ with rates $\nu_m$ and $\nu_n$, respectively. Hopping events
are assumed to be instantaneous, i.e., particle spends most of
the time being localized on a lattice site. Rate equations for
occupation numbers can be written as \cite{N13}
\begin{subequations}\label{1}
\begin{equation}
{d\hat m_i \over d\tau} = \nu_m \sum_j [\hat m_j(1-\hat m_i-\hat
  n_i)-\hat m_i (1-\hat m_j-\hat n_j)], \label{1a}
\end{equation}
\begin{equation}
{d\hat n_i \over d\tau} = \nu_n \sum_j [\hat n_j(1-\hat n_i-\hat
  m_i)-\hat n_i (1-\hat n_j-\hat m_j)]. \label{1b}
\end{equation}
\end{subequations}
Here $\hat m_i=\lbrace0,1\rbrace$ and $\hat n_i=\lbrace0,1\rbrace$ are
local occupation numbers of $m$ and $n$ particles' sort at $i$th site,
respectively, and sum is taken over nearest neighbors of $i$th site only.
 Equations (\ref{1}) are based on the assumption that it is possible
to neglect fluctuations in the number of jumps between sites $i$ and
$j$ \cite{Chumak80}, and terms of the type $\nu_m \hat m_j (1-\hat
m_i - \hat n_i)d\tau$ gives a mean number of jumps (of $m$-particle
from site $j$ to site $i$ per time $d\tau$ in this particular case).

In what follows we shall restrict our study to a macroscopic dynamics.
The simplest, if somewhat rough, way to obtain evolution equations for
the average local occupation numbers $m_i=\langle \hat m_i \rangle$
($\langle ... \rangle$ being statistical average) is to apply the mean
field approximation \cite{Leung1} or the local equilibrium
approximation \cite{Chumak80}. The latter corresponds to the
introduction of a local chemical potential associated with a given
lattice site, or a coarsened Zubarev statistical operator
\cite{Zubarev}.  In this approach we lose all the information on
fast (as compared to local equilibration time) processes and neglect
any correlations.

Next, we apply the long-wavelength approximation assuming lattice
constant $a$ to be much smaller than characteristic inhomogeneity
length scales $l_m$ and $l_n$ of $m$ and $n$ components, respectively,
$l_m\sim l_n\gg a$. In the continuum limit, equations of motion for
the densities take the form
\begin{subequations}\label{2}
\begin{equation}
{1 \over \nu_m} {dm \over d\tau} = \nabla [\nabla m - (n\nabla m -
  m\nabla n)], \label{2a}
\end{equation}
\begin{equation}
{1 \over \nu_n} {dn \over d\tau} = \nabla [\nabla n - (m\nabla n -
  n\nabla m)], \label{2b}
\end{equation}
\end{subequations}
where we have introduced dimensionless coordinate $r/a$. Obtained
equations describe smooth density profiles and can not be applied for
length scales which are comparable with a lattice constant, $l_m \sim
l_n \sim a$, where short-range correlations become significant, as near
percolation threshold, see \cite{N13}. Note that in the case where
inhomogeneity length scales of two components drastically differ, say
$l_m\gg l_n\gg a$, higher orders of space derivatives should be taken
into account in Eqs.~(\ref{2}).

More general form of Eqs.~(\ref{2}) including terms taking into
account external driving field have been obtained using mean field
approximation \cite{Leung1} or phenomenological approach \cite{S1,S2},
and exploited for investigation of phase transitions
\cite{S1,Leung1,S2}, drifting spatial structures \cite{Leung2}, and
unusual transport effects in two-component driven diffusive systems
\cite{Leung2,S3}.

The main difference of Eqs.~(\ref{2}) from the ordinary diffusion
equation is a presence of mixing flux $\bj_{mn}=n\nabla m - m \nabla n
= - \bj_{nm}$. This flux describes mutual drag of particles of one
sort by particles of another one. It is caused by the two same reasons
as a ``back correlations'' \cite{N13}, mentioned above. The first one
is a distinguishability of two different sorts of particles, e.g., by
spin or color. The second one is a local interaction (repulsion)
between particles created by excluded volume constraint.

The flows of gas components $\bj_m\equiv-\nu_m\nabla m + \nu_m(n\nabla
m - m\nabla n)$ and $\bj_n\equiv -\nu_n\nabla n + \nu_n (m\nabla n -
n\nabla m)$ can be represented as sums of diffusive and hydrodynamic
parts. For example, for $m$-component, $\bj_m=\bj_m^d+\bj_m^h$, the
flow $\bj_m^d=-\nu_m(1-n)\nabla m\equiv-D_m \nabla m$ describes
diffusion of $m$-particles through vacant sites, unoccupied by
$n$-particles, with local diffusion coefficient $D_m
\equiv\nu_m(1-n)$. Term $\bj_m^h = m(-\nu_m \nabla n)\equiv m\bV_m$
may be associated with transfer of $m$-particles by some hydrodynamic
flow with velocity $\bV_m=-\nu_m \nabla n$, where concentration $n$ of
another component plays a role of a velocity potential.
$\bj_n$ can be analyzed along the same line.

The drag of particles of one sort by particles of another sort
directly follows from Eqs.~(\ref{2}). Consider the particular case where
characteristic inhomogeneity length scale of one component is much
smaller than that of another component, say, $l_m\gg l_n\gg a$. To a first
approximation, currents of both components are governed by a
concentration gradient of the single component, $\nabla n$, at least
for a certain period of time:
\begin{equation}
{1 \over \nu_m} {dm \over d\tau} \approx \nabla (m \nabla n),\qquad {1
  \over \nu_n} {dn \over d\tau} \approx \nabla [(1-m)\nabla n].
\end{equation}
The drag effect particularly means that mass transfer of $m$-particles
can be directed along their concentration gradient, $\nabla m$, i.e.,
from region with low concentration to region with that of high one.

We will be interested in consequences of the drag effect in the
limiting case where hoping rate of $n$-particles is negligibly small
in comparison with $m$-ones, $\nu_n\ll \nu_m$. It is intuitively
clear that this is some approximation of mass transport in an
inhomogeneous medium where diffusion of mobile $m$-particles
occurs through vacant sites unoccupied by ``heavy''
$n$-particles. Considering component $n$ as a static one, we can write
reduced equation of motion for a density of mobile component $m$
\begin{equation}
{dm \over dt} = \nabla [(1-n)\nabla m + m \nabla n] = (1-n)\nabla^2m +
m\nabla^2n, \label{adv}
\end{equation}
where $t=\nu_m\tau$.  Equation (\ref{adv}) describes
advection-diffusion with compressible flow ($\nabla^2 n \not =
0$). Equation of this type is often used for the description of a
transport in various systems \cite{1,2}. In a compressible flow,
transport may exhibit intriguing effects as it is in the presence of
stable foci (traps) of the flow \cite{1}. In our case, the
hydrodynamic flow, $\sim m\nabla n$, is associated with a repulsion of
mobile particles from the frozen component. Thus, density relaxation
of mobile particles is governed by both diffusive mechanism and their
interaction with ``internal'' field $\nabla n$.

\section{\label{effects} Influence of an internal medium field}
As it is well known, the advection-diffusion can exhibit
anomalous behavior that manifests itself, in particular, in
non-classical dependence of a root-mean-square displacement on time,
$R=\langle r^2\rangle^{1/2}\sim t^\zeta$, where $\zeta \not = 1/2$ is
the exponent of anomalous diffusion \cite{1}. In the case of
compressible flow considered here ($\nabla^2 n \not = 0$), the exponent
$\zeta$ may lose universality and may depend on the degree of
compressibility. Such anomalous phenomenon is not necessarily takes
place asymptotically in the limit $t\rightarrow \infty$, but can
occur, at least, on a time scale that is less than a finite mixing
time \cite{1,B4}. On this time scale, density relaxation may strongly
depend on initial state and demonstrate peculiar behavior.

In this section we show that mass transport from low density region
towards a dense one in an inhomogeneous medium may locally accompany
subdiffusive ($\zeta<1/2$) process. Moreover, such integral
characteristic as a root-mean-square displacement $R$ does not describe
relaxation process properly. Unlike in the case of ordinary diffusion,
motion of packet front $r_f(t)$ does not coincide with
a root-mean-square displacement time dependence and has phased
behavior.

Indeed, at relatively low values of gradient $\nabla m$ mass transfer
is determined by the second term in right hand side of Eq. (\ref{adv}),
$\dot m \approx m\nabla^2 n$, which, at least over short time period,
leads to the dependence
\begin{equation}
m(r,t)\approx m(r,0)\exp (t\nabla^2 n(r)).
\end{equation}

Behavior of a density profile $m(r,t)$ in the vicinity of
frozen component distribution $n(r)$
local maxima and minima
 is different.  Near minima, where $n(r_{min}+\delta r)\approx
n_{min}(1+q^2(\delta r)^2)$, mobile particles tend to accumulate,
i.e., initial density profile is squeezing,
\begin{equation}
m(r,t)\approx m(r,0)\exp (tq^2n_{min}),\label{squeezest}
\end{equation}
which means that mass transfer may occur towards higher concentration
region. Contrary to that, mobile particles tend to evacuate from
regions close to maxima, where $n(r_{max}+\delta r)\approx
n_{max}(1-q^2(\delta r)^2)$, which can be interpreted as forcing
mobile component out by the frozen one
\begin{equation}
m(r,t)\approx m(r,0)\exp (-tq^2n_{max}).
\end{equation}
Such exponential dynamics indicates a presence of the faster process than
that of diffusive, which locally, over short time periods, leads to
the dependence $R_{loc}\sim t$, while in the case of anomalous
diffusion $R\sim t^\zeta$, with $\zeta<1$.  In this regard, such
global integral characteristic as root-mean-square displacement
\begin{equation}
R(t)=\sqrt {\langle x^2 \rangle}=\biggl({\int x^2m(x,t)\,dx\over \int
  m(x,t)\,dx}\biggr)^{1/2} \label{RMS}
\end{equation}
may be improper for the description of a diffusion process due to loss
of information on fast local dynamics.

\subsection{\label{fragm} Packet fragmentation and hoping dynamics of its front position}
In order to illustrate features of the transport phenomenon in an
inhomogeneous medium described by Eq.~(\ref{adv}), we consider
relaxation process in quasi one-dimensional case, supposing that
transverse size $L_\bot$ of a system is of the order of magnitude or
less than characteristic inhomogeneity length scale of the densities
$m$ and $n$, $L_\bot\leq l_m\sim l_n$. Then equation (\ref{adv}) reduces
to
\begin{equation}
\dot m = (1-n)\partial^2_xm + m\partial^2_xn, \label{1D}
\end{equation}
where $\partial_x$ labels one-dimensional derivative.

We will consider
spreading of initial Gaussian distribution
\begin{equation}
m(x,0)=M\exp(-x^2/4l^2)\label{gauss}
\end{equation}
in a periodic field of the frozen component
\begin{equation}
n(x)=(N/2)(1-\cos k_0x).\label{frozen}
\end{equation}
Degree of medium inhomogeneity is determined by the amplitude $N$ and
period $2\pi k_0^{-1}$ of the frozen component.

\begin{figure}
\includegraphics{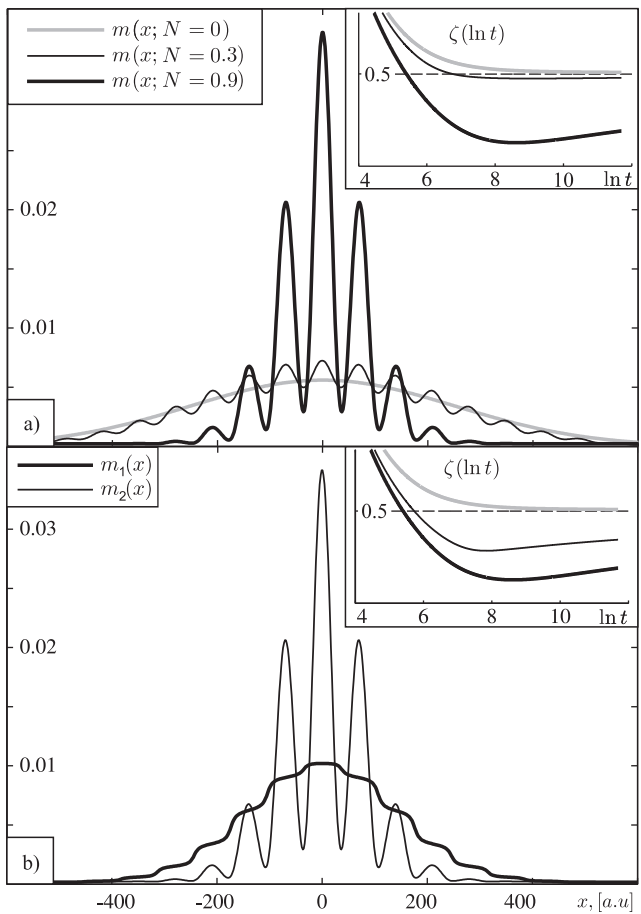}
\caption{\label{fig1} a) Density distribution $m(x,t)$ for different
  values of $N$ at $t=31888$; b) $m_1(x)$ is the density distribution,
  given by equation $\dot m_1 = \partial_x [(1-n)\partial_x m_1]$,
  $m_2(x)$ obeys Eq.~(\ref{1D}). Insets display corresponding index
  behavior $\zeta (\ln t)=\ln R(t)/\ln t$, where $R(t)$ is given by
  Eq.~(\ref{RMS}). Initial amplitude $M=0.1$, initial half-width
  $\sqrt 2 l=14.14$, and $k_0=0.1$ are used in both (a) and (b)
  cases.}
\end{figure}

As it can be seen from Fig.~\ref{fig1}(a),
presence of inhomogeneity slows down packet spreading. Such slowing
down may be caused by decreasing of effective diffusion coefficient
(renormalization of time) and/or decreasing of index $\zeta$, which
characterizes subdiffusive regime (see inset at
Fig.~\ref{fig1}(a)). However, similar behavior of index $\zeta$ (see inset at
Fig.~\ref{fig1}(b)) is given by Eq.~(\ref{1D}) with only diffusion term included $\dot
m=\partial_x [(1-n)\partial_x m]$, where relaxation has another
character, Fig.~\ref{fig1}(b). Packet fragmentation is associated with
the second term in Eq.~(\ref{adv}), which describes repulsion of light
atoms from heavy ones. If this term is dominant, mass transfer does
not go through a diffusive mechanism, described by the first term in
right hand side of Eq.~(\ref{adv}), but instead occurs due to
interaction with a heavy subsystem.

As it directly follows from Eq.~(\ref{adv}), local flow $\bj_m$ of
the mobile component is defined by two contributions,
\begin{equation}
\bj_m = -(1-n)\partial_x m- m\partial_x n. \label{flux}
\end{equation}
The first term, $\bj_m^d = -(1-n)\partial_x m\sim-(1-n)(m/l_m)$, where
$l_m$ is typical length scale of $m(x,t)$ variation near $x$,
describes standard diffuse spreading, i.e., tendency of transfer
to be directed towards concentration lowering.
Presence of the frozen component leads to a decrease of diffusion
coefficient by the factor of $1-n$. Frozen component plays the role of a
barrier for mobile particles. As concentration $n(x)$ (barrier height)
increases the velocity of penetration through the barrier by mobile
atoms (diffusive flux $\bj_m^d$) decrease as $1-n$. The flux
$\bj_m^h  = - m\partial_x n\sim- m(nk_0)$ describes repulsion of
mobile atoms from frozen one. Fluxes $\bj_m^d$ and $\bj_m^h$, defined
by a density gradients $\partial_x m$ and $\partial_x n$, may have
different signs and behave as two competing fluxes. If $1/n-1<l_mk_0$,
then total flux $\bj_m$ may be locally directed towards region with
higher concentration, which means local compression of mobile atom
distribution.

\begin{figure}
\includegraphics{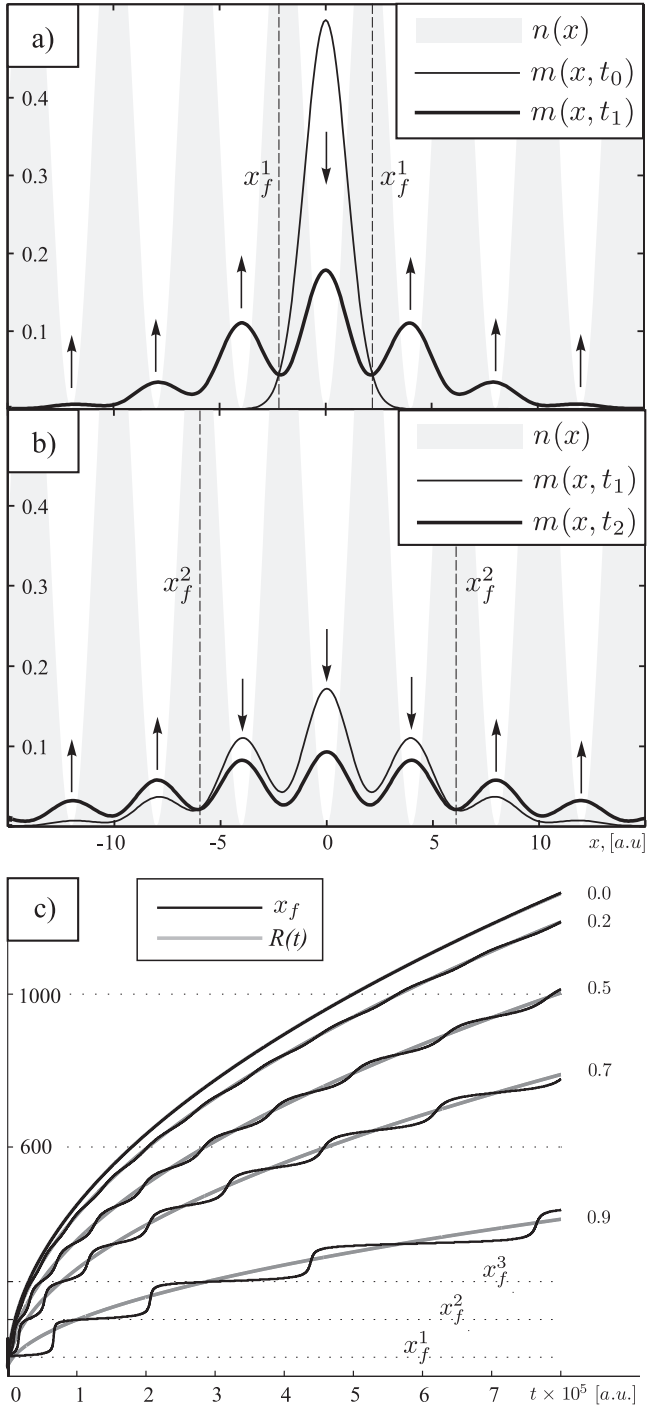}
\caption{\label{fig2} Packet dynamics during time intervals:
  a)~[$t_0,t_1$], b)~[$t_1, t_2$];
  ($t_0=0,\ t_1=14\times10^3,\ t_2=45\times10^3$ time units). Initial
  amplitude $M=0.5$, amplitude of the frozen component $N=0.7$. Shaded
pattern shows density distribution profile of the frozen component $n(x)$,
given by Eq.~(\ref{frozen}), arrows
  denote whether sub-packet is squeezing (growing) ($\uparrow $) or
  spreading ($\downarrow$) on certain time interval. Dashed lines
  $x_f^i$ denote region of main localization during corresponding time
  interval.\\ c) Time dependence of packet front $x_f(t)$ defined as a
  point where $\dot m=0$, and a root-mean-square displacement $R(t)$
  defined by Eq.~(\ref{RMS}), for different values of $N$. Lines'
  labels correspond to values of $N$. Dotted lines $x_f^i$ correspond
  to the dotted lines at Figs.~\ref{fig2} (a) and (b).}
\end{figure}

Figure \ref{fig2} illustrates dynamics of initial Gaussian
distribution spreading, which has phased nature. Front propagation
through the first barrier towards neighboring local profile minimum
(depicted at Fig.~\ref{fig2}(a)) consists of two stages. At first,
particles locally accumulate in the minimum due to $\bj_m^h$ domination
until the condition $2/N-1>l_mk_0$ for $m(x,t)$ profile is met, or
in other words, until local ``pressure'' ($\propto \partial_x m$) gets
high enough to overcome repulsion from next barrier. Expression
$2/N-1>l_mk_0$ defines the condition for diffusive penetration through
the barrier. Then process repeats for the next local minimum,
while in the previous one particles undergo usual spreading,
Fig.~\ref{fig2}(b). Such staged process describes motion of a packet
front $x_f(t)$, which position is determined by condition $\dot
m(x_f,t)=0$, see Fig.~\ref{fig2}(c).

Front $x_f(t)$ divides $x$-axis in two regions: (i) $|x|<x_f$, where
concentration $m(x,t)$ is decreasing for any $x$ and density
relaxation is going mainly due to diffusion with diffusion coefficient
renormalized by a profile of the frozen component; (ii) $|x|>x_f$, where
concentration is increasing and relaxation is strongly affected by
a repulsion from the frozen component (medium inhomogeneity). The latter
leads to a local compression and density increase.

As can be seen from Fig.~\ref{fig2}(c), medium inhomogeneity, which in
our case is determined by an amplitude $N$, leads to appearing of difference between packet's front motion $x_f(t)$ and
root-mean-square displacement $R(t)$ dynamics. Front $x_f(t)$ exhibits
phased behavior and is governed by two linear in time, fast and slow
processes. Packet front spends most of the time being localized inside
a barrier and performs quick jumps into neighboring barriers while
root-mean-square displacement (Eq.~(\ref{RMS})) gives only an averaged
dynamics of front motion, see Fig.~\ref{fig2}(c).

\subsection{\label{compr} Compressibility of initial density distribution}
\begin{figure}
\includegraphics{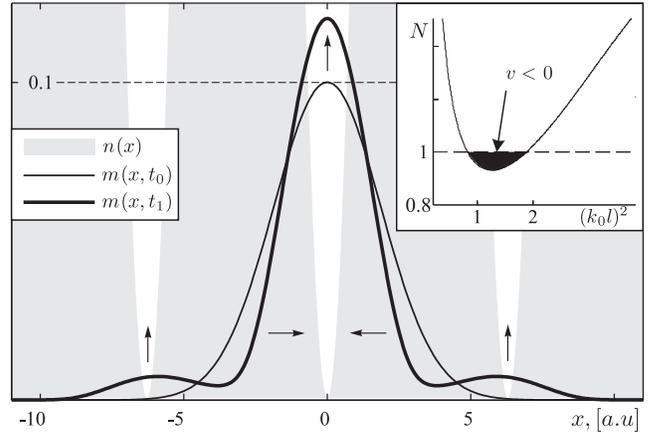}
\caption{\label{fig3} Central peak squeezing (growth) of a mobile
  component density profile during initial time interval [$t_0,t_1$],
  ($t_0=0,\ t_1=3\times10^4$ time units). Initial amplitude $M=0.1$,
  $N=0.9$, initial half-width $\sqrt 2 l=1.7$, and $k_0=1.05$. Shaded
pattern shows density distribution profile of the frozen component $n(x)$,
given by Eq.~(\ref{frozen}), arrows
  indicate directions of packet evolution. Filled region at the inset is
  determined by inequality (\ref{ineq}) and condition $N<1$, and
  corresponds to a negative velocity of packet motion.}
\end{figure}

Compression (growth) of density distribution also can occur for
the central packet during certain initial time interval. Indeed, from
Eqs.~(\ref{adv}),(\ref{gauss}), and (\ref{frozen}) it is easy to
estimate behavior of distribution amplitude at the peak over short
time periods,
\begin{equation}
m(0,t)\approx M+\frac{Mt}{2l^2}(N k_0^2 l^2-1).
\end{equation}
Amplitude $m(0,t)$ tends to increase if $N>(k_0l)^{-2}$. On the other
hand, density distribution compression means that velocity (time
derivative) of a mean-square displacement $v(t)=d\langle
x^2\rangle/dt$ is negative. The latter may be roughly estimated in
Fourier domain
\begin{equation}
v(t)=-{1\over m_0}\partial^2_k\dot m_k\big |_{k=0},
\end{equation}
where $m_k(t)$ is the Fourier image of $m(x,t)$ and obeys equation
\begin{eqnarray}
\dot m_k=-(N/4)[(k^2-2kk_0)m_{k-k_0}]-{}\quad{}\label{advfourier}
\nonumber \\ -(N/4)[(k^2+2kk_0)m_{k+k_0}]-k^2[1-(N/2)]m_k.
\end{eqnarray}
Since $v(t)$ is determined by limit $k\to 0$, it is reasonable to
approximate Eq.~(\ref{advfourier}) by its expansion for small
$k$ ($k\ll k_0$)
\begin{eqnarray}
\dot m_k \approx -k^2[1-(N/2)]m_k-(N/2)k^2m_{k_0} \nonumber
\\ -Nk^2k_0\frac{\partial}{\partial
  k_0}m_{k_0}-(1/4)Nk^4\frac{\partial^2}{\partial k_0^2}m_{k_0},\quad
\end{eqnarray}
where we have omitted terms with $m_{2k_0}\approx0$ assuming that
typical scale of inhomogeneity for $m(x,t)$ is of the order or less
than that of for $n(x)$, i.e., $k\leq k_0$. Taking limit $k \to 0$ we obtain
expression for velocity
\begin{equation}
v\approx(2-N)-\frac{N}{m_0}\biggl(1+2k_0\frac{\partial}{\partial
  k_0}\biggr)m_{k_0}\label{veloc}.
\end{equation}
Here $m_{k_0}$ is defined by the equation
\begin{equation}
\dot
m_{k_0}\approx-k_0^2\biggl(1-\frac{N}{2}\biggr)m_{k_0}+\frac{N}{4}m_0.
\end{equation}
From (\ref{veloc}) condition for packet squeezing ($v<0$) directly
follows, see inset at Fig.~\ref{fig3},

\begin{equation}\label{ineq}
2/N<1+(4k_0^2l^2-1)\exp (-k_0^2l^2).
\end{equation}

Numerical solution of Eq.~(\ref{1D}) corresponding to such a case is
shown at Figure~\ref{fig3}. Initial compression lasts for certain time
period after which diffusion goes by means of sequential overcoming of
potential barriers (as it was discussed in previous subsection). Note
that the estimate (\ref{ineq}) is rough and initial packet squeezing
actually appear at smaller values of amplitude $N$, as shown at
Fig.~\ref{fig3}.

\section{\label{station} The Stationary case}
As it was shown in the previous section, interaction of diffusing
particles with inhomogeneities of a medium leads to a local particle
accumulation, i.e., local mass transport towards higher
concentration. It is interesting that such transport effect can be
realized at global scale, throughout a whole system, with arbitrary
distribution $n(x)$.  In order to illustrate such a possibility we consider quasi
one-dimensional boundary problem
\begin{equation}
(1-n)\partial_x^2m+m\partial_x^2n=0 \label{stat}
\end{equation}
with boundary conditions
\begin{equation}
m(0)=m(0),\quad m(L)=m(L),
\end{equation}
where $L$ is a sample length.  Solution of Eq.~(\ref{stat}) has the
form
\begin{equation}
m(x)=(1-n(x))\biggl({m(0) \over 1-n(0)}- J \int _0^x {d\xi\over
  (1-n(\xi))^2}\biggr),
\end{equation}
where $J$ is a total particle flux through the system
\begin{eqnarray}
J=-\biggl (\int_0^L \frac{d\xi}{[1-n(\xi)]^2}\biggr)^{-1}\biggl (
{m(L)\over 1-n(L)}-{m(0)\over 1-n(0)} \biggr)= \label{J}
\nonumber\\ ={1 \over L} \int _0^L \bj_m(x)\ dx. \quad \quad \quad
\quad
\end{eqnarray}
In the last expression $\bj_m(x)$ is given by Eq.~(\ref{flux}).

As we see, stationary density distribution $m(x)$ may be inhomogeneous
even if concentrations at the boundaries $m(L)=m(0)$ are equal.

As it was mentioned above, mass transfer in such a system is governed
not only by the mean field of $m(x)$ gradient, but also by interaction
of a mobile subsystem with a frozen one. As it follows directly from
Eq.~(\ref{flux}) mass transfer may go from lower concentrations
towards higher one. Assuming $m(L)>m(0)$ one gets, that flux $J$ is
directed towards the higher concentration region, i.e., from the
boundary $x=0$ to the boundary $x=L$, if the following condition is
satisfied
\begin{equation}
{m(L)\over m(0)}<{1-n(L)\over 1-n(0)}.
\end{equation}
Thus total flux direction is determined by boundary conditions for $m$
and $n$ components.

\section{\label{concl} Summary and Discussion}
The presence of a second sort of particles in lattice gas leads to
peculiar transport effects.  In the case of a two-component gas
mass transport is affected by an action of an additional flux, contrary
to the case of a single component gas of indistinguishable
particles. This flux is associated with mixing of different gas
components (interdiffusion term) and in a presence of interaction
between particles such as excluded volume constraint in our case, and may
lead to a drag of one sort of particles by another one. If particles
of one sort have negligibly small mobility in comparison with another
sort, so that one sort is assumed to be static, then the mixing flux,
or its part, of mobile particles transforms into a stationary
hydrodynamic flow that may drag mobile particles. As a result, density
relaxation of mobile particles occurs via diffusion process and/or
via an interaction with a medium, i.e., with a frozen gas component.

In our paper we have presented a case of a system where diffusive mechanism
cooperates with density relaxation under action of ``internal medium
field'', given by a periodic distribution of a static component. Such
mutual cooperation leads to unusual transport with local mass transfer
resulting in an increase of concentration.

Indeed, as it was shown, if medium is strongly inhomogeneous, its
``field'' leads to a local accumulation of diffusing particles in minima
of inhomogeneity profile. Presence of such a local processes is
manifested by fragmentation of a packet during its spreading. The
dynamics of a diffusive process also becomes peculiar and mass
transport occurs as sequential penetration through inhomogeneity
barriers, i.e. regions with high concentration of the static
component. This leads to a phased character of a packet front motion
that consists of a sequence of fast and slow processes (moves) with
linear in time coordinate dependence. Such front behavior is similar
to motion of a defect (slow motion inside barrier and fast passing
through inhomogeneity minima). Contrary to the case of ordinary
diffusion behavior of a root-mean-square displacement $R(t)$ and
that of a packet front $x_f(t)$ do not coincide. $R(t)$ reflects only an
averaged packet front dynamics. Local accumulation of particles, which
accompanies diffusive process and sequential barriers overcoming, also
slows down packet spreading. The latter entails subdiffusive regime.
In stationary case interaction of diffusing particles with medium
inhomogeneity can lead to mass transport directed from
a boundary with low concentration to a boundary with that of high one.

Note, that quasi one-dimensional case considered in this work is
rather illustrative, because dimensionality reduction leads to
enhancement of particle-particle correlations. In two- and
three-dimensional cases demonstrated effects may not be so
pronounced. However, accounting for interparticle interaction on
nearest neighboring sites leads to essential changes in mass transport
\cite{N11Chum} and may enhance discussed effects.

Note, that assigning appropriate mobilities to components in
Eqs.~(\ref{2}) we can model not only inhomogeneous media but
also systems with quenched randomness, as it was pointed out in
\cite{S2,S1}, or, for example, quenched inhomogeneities, dynamically
generated in glasses \cite{glass}.

Distinguishability of two sorts of particles and excluded volume
constraint are responsible for unusual transport effects appearing in
two-component lattice gas. Note, these effects may speculatively be
considered as a purely statistical consequence of particles'
distinguishability in a multicomponent non-interacting Fermi gas.

Equations (\ref{2}) have been obtained using rough approximations,
which entails certain limitations on their validity. First of all we
neglect fast processes in the system, as it is generally the
case for the diffusion approximation, the framework basic
Eqs.~(\ref{1}) are written in. In addition, Langevin source of
fluctuations in a number of jumps between lattice sites was omitted
in Eqs.~(\ref{1}). We also loose information on fast processes
applying the mean-field approximation which is equivalent to the local
equilibrium one. This approximation also means that we neglect any
correlations in the system, in particular, short-range ``back
correlations'' (local memory effect) \cite{N13, Chumak2002} that are
known to contribute to effective diffusive process. The
long-wavelength approximation does not allow to consider mass
transport on length scales comparable with the lattice distance. In
particular, it is in a vicinity of percolation threshold where
short-range correlations become significant.

\begin{acknowledgments}
Authors thank P.~M.~Tomchuk, B.~I.~Lev, V.~V.~Ieliseieva, O.~O.~Chumak,
K.~S.~Karplyuk, V.~V.~Obukhovsky, and D. A. Bevzenko for
stimulating discussions and useful suggestions.
\end{acknowledgments}

\end{document}